\documentclass[onecolumn,showpacs,preprintnumbers,amsmath,amssymb]{revtex4}
 \usepackage{graphicx}
\usepackage{latexsym}  
\usepackage{dcolumn}
\usepackage{graphicx, subfigure}

\usepackage{graphicx}
\usepackage{dcolumn}
\usepackage{bm}

\usepackage{bm}

\begin{document}

\newcommand{\eq}{\begin{equation}}                                                                         
\newcommand{\eqe}{\end{equation}}             

\title{Laser assisted proton collision on light nuclei at moderate energies}

\author{ I. F. Barna and S. Varr\'o}
\address{Wigner Research Centre of the Hungarian Academy 
of Sciences, \\ Konkoly Thege \'ut 29 - 33, 1121 Budapest,  Hungary\\ 
and \\ 
ELI-HU Nonprofit Kft.,  Dugonics T\'er 13, 6720 Szeged, Hungary}
\date{\today}

\begin{abstract}
We present analytic angular differential cross section model for laser assisted proton 
nucleon scattering on a Woods-Saxon optical potential where the nth-order photon 
absorption is taken into account simultaneously. 
As a physical example we calculate cross sections for proton - $^{12}$C collision at 49 MeV  in the laboratory frame where the laser intensity is in the  range of  $ 10^{7} - 10^{21}$ 
 W/cm$^2$ at optical frequencies. The upper intensity limit is 
slightly below the relativistic regime.  
\end{abstract}

\maketitle

\section{Introduction}
Nowadays optical laser intensities exceeded the $10^{22}$ W/cm$^2$ limit where radiation effects dominate the electron dynamics. In the  field of laser-matter interaction 
a large number of non-linear response of atoms, molecules and plasmas can be investigated 
both theoretically and experimentally.  Such interesting high-field phenomena are high harmonic generations, or plasma-based laser-electron acceleration. 
 This field intensities open the door to high field quantum electrodynamics phenomena like vacuum-polarization effects of pair production  \cite{muller}.  
In most of the presented studies the dynamics of the participating electrons are investigated.    
Numerous surveys on laser assisted electron collisions are available as well \cite{kam}.
However, there are only few nuclear photo-excitation investigations done where some low-lying first excited states of medium of heavy elements are 
populated with the help of x-ray free electron laser pulses \cite{gunst}.  Nuclear excitation by atomic electron re-scattering in a laser field was investigated as well \cite{korn}. 
 Various additional concepts are under consideration for photo-nuclear reactions 
by laser-driven gamma beams \cite{habs}. 

To our knowledge there are no publications available where laser assisted proton nucleus collisions  (or radiative proton-nucleus scattering)
were investigated.  This is the  goal of our recent paper. 
We consider the full optical potential of Woods-Saxon(WS) type \cite{woods} with the proper 
parametrization for moderate energy proton - $^{12}$C collision \cite{abd}. 
The optical potential formalism has been a very successful method to study the single particle 
spectra of nucleus in the last  four decades. Detailed description of the validity of this formalism 
can be found in a nuclear physics textbooks or in monographs like \cite{grei, ger}. 
 
The nuclear physics community recently managed to evaluate the closed analytic form of the 
Fourier transformed WS interaction \cite{hlophe} which is a great success.   
We incorporate these results into a first Born approximation scattering cross section formula where the initial and final proton wave functions are Volkov waves 
and the induced photon emission and absorption processes are taken into account up to arbitrary orders \cite{bunk, bunk1,faisal,gont,bergou,kroll,faisal2}.
We end our study with a physical example where 49 MeV protons are scattered on  $^{12}C$ nuclei in the field of a  Titan sapphire laser.  

\section{Theory} 
In the following we summarize our applied  non-relativistic quantum mechanical description.  
The laser field is handled in the classical way via the minimal coupling. The laser beam is taken to be linearly polarized and the dipole approximation is used. 
If the dimensionless intensity parameter (or the normalised vector potential)  $a_0 = 8.55 \cdot 10^{-10} \sqrt{ I (\frac{W}{cm^2})} \lambda (\mu m)$ of the laser field is smaller than unity 
the non-relativistic description in dipole approximation is valid. For 800 nm laser wavelength this means a critical intensity of $ I = 2.13 \cdot 10^{18} $W/cm$^2$.  
In case of protons $a_0$ is replaced by $a_p=((m_p/m_e)^{-1})a_0$, where the proton to electron mass ratio is $(m_p/m_e)$=1836 \cite{nist}). 
Accordingly, for 800 nm wavelength the critical intensity for protons is $I_{crit} = 3.91 \cdot 10^{21}$ W/cm$^2$.  
Additionally, we consider moderate proton kinetic energy, not so much above the Coulomb barrier and neglect the interchange term between the proton projectile an the 
target carbon protons.  This proton exchange effect could be included in the presented model with the help of Woods-Saxon potentials  of non-local type \cite{imre} but  not in the scope 
of the recent study.

To describe  the scattering process of a proton on a nucleus in a spherically symmetric field  
the following Schr\"odinger equation has to be solved,  
\begin{equation}
\left[ \frac{1}{2m} \left(  {\bf{p}}- \frac{e}{c} {\bf{A}} \right)^2 + U(r) \right]\Psi  = i \hbar  \frac{\partial \Psi}{\partial t}, 
\end{equation}
where ${\bf{p}}, \Psi $  are the momentum operator and the  wave function of the proton, ${\bf{A}}(t)= A_0 {\bf{\cal{\epsilon}}} cos(\omega t)$ is the vector potential of the external laser field with unit polarisation vector ${\bf{\cal{\epsilon}}} $ and $U(r)$ is the scattering potential. 
We search the solution in the following form of 
\begin{equation}
 \Psi = \varphi_{p_i} + \Psi_c, 
\end{equation}

where 
\begin{equation}
\varphi_p({\bf{r}},t) =  \frac{1}{(2\pi \hbar)^{3/2}} e^{\frac{i}{\hbar}({\bf{p}}{\bf{r}}- Et  ) } exp\left[ 
-\frac{i}{\hbar}\frac{p^2 t}{2m } - \int_0^t dt' \left( \frac{e{\bf{p}} {\bf{A}}(t') }{mc} + \frac{1}{2m} \left( \frac{eA(t')}{c} \right)^2 \right )
\right] 
\end{equation}
is the Volkov wave function of the scattering proton.
 We note at this point that the $A^2$ term drops out from the transition amplitude, because it depends only on time (and does not depend on the proton's momentum).
 The correction term is 
\begin{equation}
\Psi_c = \int d ^3  p a_{{\bf{p}}}(t) \varphi_p,  
\end{equation}
where $ a_{{\bf{p}}}(t)$ is the scattering amplitude. 
Calculating the transition matrix element between the final and the initial states in the
first Born approximation for $a_{{\bf{p}}_f}(t)$ at $   t \rightarrow \infty$ we get 
\begin{equation}
T_{fi} =  \langle   \varphi_f   |U| \Psi_i  \rangle = -\frac{ U({\bf{q}})}{(2\pi \hbar)^3} \sum_{n=-\infty}^{\infty}  2\pi i \delta \left(   \frac{p_f^2-p_i^2}{2m} + n \hbar \omega  \right) J_n(z), 
\end{equation}
where  $U({\bf{q}})$ is the Fourier transformed of the scattering potential with the momentum transfer of 
 ${\bf{q}}  \equiv {\bf{p}}_i  - {\bf{p}}_f  $ where  ${\bf{p}}_i$ is the initial and 
${\bf{p}}_f$ is the final proton momenta.    The absolute value is $    q = \sqrt{p_i^2 + p_f^2 - 2p_ip_f cos(\theta_{p_i,p_f})}$. 
In our case, for 49 MeV energy protons absorbing optical photons the following approximation is valid 
$q \approx 2 p_i  |sin(\theta/2)| $.

 The Dirac delta describes photon absorptions $(n<0)$ and emissions $(n>0)$.  
$J_n(z)$ is the Bessel function with the argument of  
\begin{equation}
z   \equiv  \frac{m_e}{m_p} a_0 ( {\bf{\hat{ q}}} {\bf{\cal{\epsilon}}}) \frac{2p_i}{\hbar k_0} |sin(\theta/2)|  	
\end{equation}  
where  $m_e, m_p$ are the electron and proton masses, $a_0$ is the dimensionless intensity parameter, (given above),  $   {\bf{\hat{ q}}}  $ and   $ {\bf{\cal{\epsilon}}}  $ are the unit vectors of the  
momentum transfer and the laser polarisation direction.    
It can be shown with geometrical means that for low energy photons where ($E_{ph} < E_{p^+}$) the angle in the scalar product of $\hat{{\bf{q}}}  {\bf{\cal{\epsilon}}} \equiv cos \chi$ 
is $\chi = \pi/2 - \theta/2$ where $\theta$ is the scattering angle of the proton varying from 0 to $\pi$.    \\ 
From $ \frac{p_i}{\hbar k_0} = \sqrt{\frac{m_p}{m_e}}  \sqrt{\frac{2 m_e c^2 E_i}{\hbar^2 \omega_0^2}}$ 
collecting the constants together the final formula for z reads
\begin{equation} 
z =  \frac{1.4166\times 10^{-3} }{\hbar \omega_0}   \sqrt{\frac{E_p}{1836}}  \sqrt{I}  \times cos(\chi) \times | sin(\theta/2) |, 
\end{equation} 
where the laser energy $\hbar \omega_0$ is measured in eV, the proton energy $E_p$  in MeV and the laser intensity I in W/cm$^2$. 
 
The final differential cross section formula for the laser associated collision with simultaneous nth-order photon absorption and emission processes is 
\begin{equation}
\frac{d \sigma^{(n)}}{d \Omega} = \frac{p_f}{p_i}J_n^2(z) \frac{d \sigma_B}{d \Omega}. 
\label{cross}
\end{equation} 
The $ \frac{d \sigma_B}{d \Omega} = \left(\frac{m}{2\pi \hbar^2} \right)^2 |U({\bf{q}})|^2$ 
is the usual Born cross section for the scattering on the potential U(r) alone (without the laser field). 
The expression Eq. (\ref{cross}) was calculated with different authors using different methods \cite{bunk, bunk1,faisal,gont,bergou,kroll,faisal2}. \\ 

The scattering interaction $U(r)$  is the sum of the Coulomb potential of a uniform charged sphere \cite{rudchik} and a short range optical Woods-Saxon potential \cite{woods}
\begin{equation}
U(r) = V_c(r) + V_{ws}(r) + i[W(r) + W_s(r)] + V_{ls}(r) {\bf{l}} \cdot {\bf{\sigma}} 
\end{equation} 
where the Coulomb  term is 
\begin{eqnarray}
V_c &=& \frac{Z_p Z_t e^2}{2R_0} \left(3 - \frac{r^2}{R_c^2} \right) \hspace*{1cm} r < R_c  \nonumber  \\
V_c &=&\frac{Z_pZ_t e^2}{r}  \hspace*{2.8cm} r \ge R_c  
\end{eqnarray}
where $R_c = r_0 A_t^{1/3}$ is the target radius calculated from the mass number of the  target 
with $r_0 = 1.25 fm$.   $Z_p, Z_t$ are  the charge of the projectile and the target and $e$ is the elementary charge. 
This kind of regularised Coulomb potential helps us to avoid singular cross sections and routinely used in nuclear physics. 

The short range nuclear part  is given via 
\begin{eqnarray}
V(r) &=& - V_r f_{ws}(r,R_0,a_0) \nonumber \\ 
W(r) &=& - V_v f_{ws}(r,R_s,a_s) \nonumber \\ 
W_s(r) &=& - W_s(-4a_s) f_{ws}'(r,R_s,a_s) \nonumber \\
 V_{ls}(r) &=& -(V_{so} + iW_{so})(-2) g_{ws}(r,R_{so},a_{so}) \nonumber \\ 
f_{ws}(r,R,a) &=& \frac{1}{1+ exp \left( \frac{r-R}{a} \right)} \nonumber \\ 
f'_{ws} (r,R,a)&=& \frac{d}{dr} f_{ws}(r,R,a) \nonumber \\   
g_{ws}(r,R,a) &=& f'_{ws}(r,R,a)/r.
\end{eqnarray}
The constants $V_r, W_v,V_{so} $ and $ W_{so}$ are the strength parameters, 
and $a_{0,s,so}, R_{0,s,so}$ are the diffuseness and the radius parameters given for large number of nuclei. 
As we will see at moderate collisions energies the complex tems become zero. 
In the last part of the present paper we will use the numerical  parameters of \cite{abd} for proton carbon-collision. 

According to the work of Hlophe \cite{hlophe} the complete analytic form of the Fourier transform of the Woods-Saxon potential is available 
\begin{eqnarray} 
V_s(q) =  \frac{V_r}{\pi^2} \left\{ \frac{\pi a_0 e^{-\pi a_0 q}}{q(1-e^{-2\pi a_0 q})^2} \left[ 
R_0(1-e^{-2\pi a_0 q})cos(qR_0)-  \pi a_0 (1+ e^{-2\pi a_0 q}) sin(qR_0) 
\right]  -  \right. \nonumber \\ 
 \left. a_0^3 e^{-\frac{R_0}{a_0}} \left[ \frac{1}{(1+a_0^2q^2)^2} - \frac{2e^{-\frac{R_0}{a_0}}}{(4+
a_0^2 q^2)^2}  \right]    \right\} .  
\label{ws}
\end{eqnarray}
For the $W(q)$ imaginary term, the same expression was derived with $W_v,a_s,R_s$ instead of $V_r,a_0$ and $R_0$.
The surface term $W_s(r)$ gives the following formula in the momentum space: 
\begin{eqnarray}
W_s(q) = - 4 a_s \frac{W_s}{\pi^2} \left\{ \frac{\pi a_s e^{-\pi a_s q}}{(1-e^{-2\pi a_s q})^2} \left[ 
(\pi a_s(1+e^{-2\pi a_s q})-  \frac{1}{q} (1- e^{-2\pi a_s q}) )cos(qR_s) 
   + R_s(1-e^{-2\pi a_s q})sin(qR_s)  \right] \right. \nonumber \\  +
 \left. a_s^2 e^{-\frac{R_s}{a_s}} \left[ \frac{1}{(1+a_s^2q^2)^2} - \frac{4e^{-\frac{R_s}{a_s}}}{(4+
a_s^2 q^2)^2}  \right]    \right\}.    
\label{s}
\end{eqnarray}
The transformed spin-orbit coupling term leads to  
\begin{eqnarray}
V_{ls}(q) =    -\frac{a_{so}}{\pi^2}(V_{so}+ iW_{so})
 \left\{   \frac{2 \pi e^{-\pi a_{so}q}}{1-e^{-2\pi a_{so}q}} sin(q R_{so})  + 
e^{\frac{-R_{so}}{a_{so}}} \left(  \frac{1}{1+a_{so}^2 q^2} -   \frac{2e^{-R_{so}} {}}{4+a_{so}^2 q^2}\right)
  \right\}.
\label{ls}
\end{eqnarray}
where  the momentum transfer is defined  as above  $ {\bf{q}} \equiv  {\bf{p_i}} - {\bf{p_f}}. $ 
The  low energy transfer approximation formula $q \approx 2 p_i  |sin(\theta/2)| $ is valid. 

The Fourier transform of the charged sphere Coulomb field is also far from being trivial
\begin{eqnarray}
V_c(q) &=&    \frac{Z_pZ_t e^2}{2^{\frac{5}{6}} \sqrt{\pi} q^3} \left( -2\cdot 3^{\frac{1}{3}} q  \cos[2^{\frac{2}{3}} 3^{\frac{1}{3}} q]  + 
  2^{\frac{1}{3}} (1 + 2\cdot 2^{\frac{1}{3}} 3^{\frac{2}{3}} q^2) \sin[
    2^{\frac{2}{3}} 3^{\frac{1}{3}} q] \right)  +     \nonumber  \\
 &&3Z_pZ_t e^2 \sqrt{\frac{2}{\pi}} \left(   \frac{i \pi|q|  }{2q}  - \mathrm{Ci} [2^{\frac{2}{3}}  3^{\frac{1}{3}} q] + \log(q) - \log|q| -  i\mathrm{Si} [2^{\frac{2}{3}}  3^{\frac{1}{3}} q] \right)    
\end{eqnarray}
where Ci and Si  are the cosine integral and the sine integral functions, respectively. 

\section{Results} 

We applied the outlined method to 49 MeV proton - $^{12}$C scattering. 
Table I contains the parameters of the used Woods-Saxon potential. 

\begin{center}
\begin{tabular} {|c |c|}  \hline \hline 
Name of the parameter & Numerical value    \\    \hline
$V_R$  &  31.31  (MeV)  \\
$R_0 $ & 1.276 (fm) \\ 
$a_0$ & 0 .68 (fm)  \\   \hline
$W_s$  &  5.98  (MeV)  \\
$r_s $ & 0.890 (fm) \\ 
$a_s$ &  0.586 (fm)  \\   \hline 
$V_{so}$  &  2.79 (MeV)  \\
$R_{so} $ & 0.716 (fm) \\ 
$a_{so}$ & 0 .222 (fm)  \\   \hline \hline
\end{tabular}
\center{ Table 1. The parameters of the applied potential for proton - $^{12}C$ collision at $E_i = 49$ MeV. \\  Note, that the 
complex part  $W_v$ and the the complex part of the spin-orbit term $W_{so}$ are zero at this energy. }
\end{center}

Figure 1 presents the angular differential cross section in the first Born approximation of the various Woods-Saxon potential terms for 49 MeV elastic 
proton -  $^{12}$C scattering. The different lines represent the different terms (\ref{ws},\ref{s},\ref{ls}). 
The laboratory frame is used in the calculation.  For a better transparency the contributions of the regularised Coulomb term is not presented.  
   Our calculated total cross section of the elastic scattering is  201 mbarn which is
 consistent with the data of \cite{abd}.

\begin{figure}
\scalebox{0.5}{
\rotatebox{0}{\includegraphics{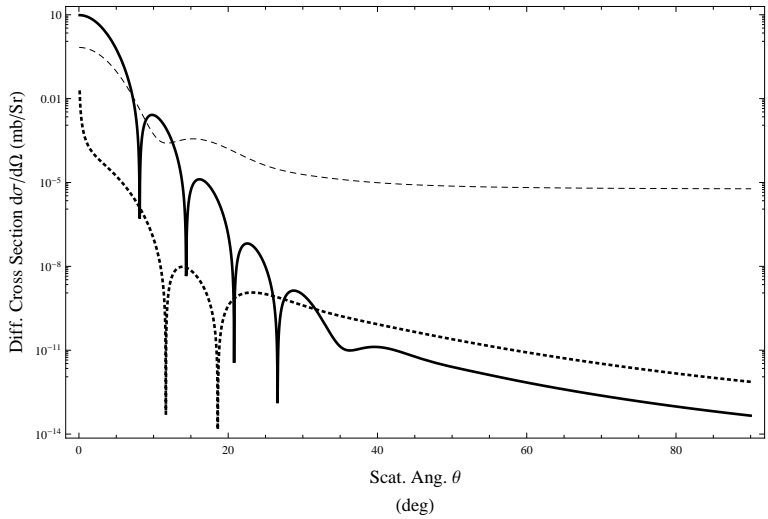}}}  
\caption{The angular differential cross sections in the first Born approximation of the various  Woods-Saxon potentials tems for 49 MeV elastic 
proton -  $^{12}$C scattering. The solid, dashed, and dotted lines are the contributions  of  Eqs. 
(\ref{ws},\ref{s},\ref{ls}), respectively. Note, the different smoothness and different back scattering values of the different terms.}  	
\label{egyes}       
\end{figure}
 	
\begin{figure}
\scalebox{0.5}{
\rotatebox{0}{\includegraphics{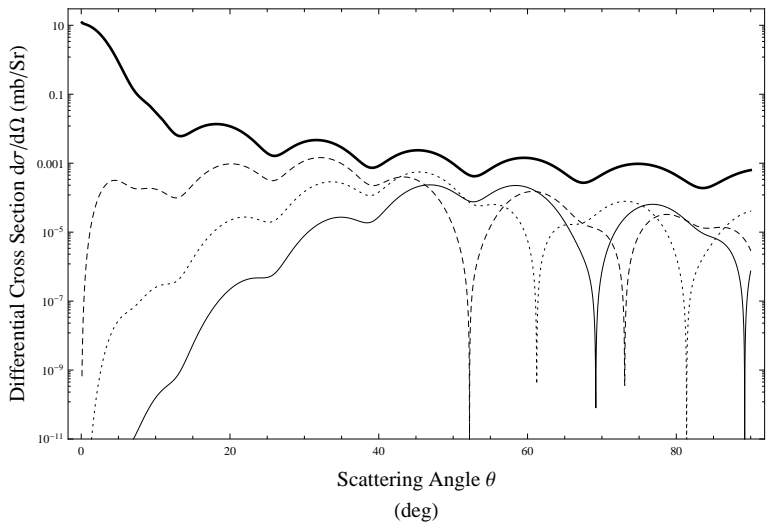}}}  
\caption{The calculated angular differential cross sections from Eq. (8) for  $I = 4.56 \times 10^7$  W/cm$^2 $ laser field intensity (${\cal{I}} = 1$). 
 The thick solid, thin long dashed, thin short dashed and thin solid lines are for $ n = 0,1,2,3 $ 
photon absorptions, respectively.}  	
\label{kettes}       
\end{figure}

In our case the laser photon energy is  $\hbar \omega_0 = 1.56 $ eV, which means 800 nm wavelength and  the proton energy is 
$E_i = 49$ MeV.  With these values the argument of the Bessel function  Eq. (7) becomes 
the following 
\begin{eqnarray}
z = 1.48\times 10^{-3}  \sqrt{I} \times cos(\chi) \times | sin(\theta/2) | = {\cal{I}} \times cos(\chi) \times | sin(\theta/2) | . 
\end{eqnarray} 
 Figure \ref{kettes} presents the angular differential cross section for $n=0,1,2,3$ photon absorptions for  $ {\cal{I}} = 1$ 
which means $I = 4.56\times 10^7$  W/cm$^2$ intensity.  The single, double and triple photon absorption total cross sections are 
2.26, 1.1 and 0.71 mbarn, respectively. 

\begin{figure}
\scalebox{0.5}{
\rotatebox{0}{\includegraphics{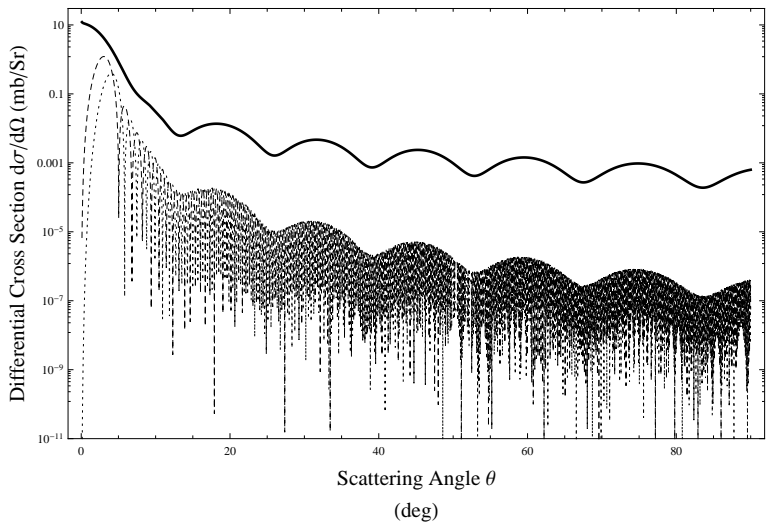}}}  
\caption{  The calculated angular differential cross sections from Eq. (8) for $I = 4.56 \times 10^{11}$  W/cm$^2 $ laser field intensity (  ${\cal{I}}=100$).  
  The thick solid, thin long dashed, thin short dashed are for $ n = 0,1,2$ 
photon absorptions.}    	
\label{harmas}       
\end{figure}

Figure \ref{harmas} shows  the angular differential cross section for $n=0,1,2$ photon absorptions for $ {\cal{I}} = 100$ 
which means $I = 4.56\times 10^{11}$  W/cm$^2$ intensity. Note, that the cross sections for single and double photon absorption 
are almost the same.  The single photon absorption total cross section is 0.5 mbarn. 

\begin{figure}
\scalebox{0.5}{
\rotatebox{0}{\includegraphics{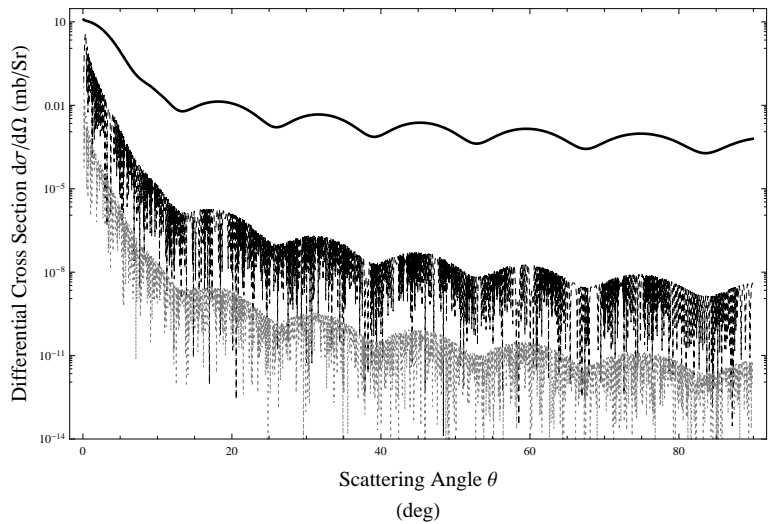}}}  
\caption{  The calculated angular differential cross sections for $n=0$ and photon absorption. The black thin solid line is for  $I = 4.56 \times 10^{15}$  W/cm$^2 $ 
intensity ( ${\cal{I}}=10000$).  The dashed gray line is close to the relativistic threshold  with $I = 2.0 \times 10^{21}$  W/cm$^2 $ laser field intensity (${\cal{I}}=6.61 \times 10^6). $ }    	
\label{negyes}       
\end{figure}

For large laser field intensities, which means large z arguments of the Bessel functions  the following asymptotic expansion can be used for a fixed index \cite{abr}  
\begin{eqnarray}
J_n(z) = \sqrt{2/(\pi z)} Cos(z-n \pi /2 -\pi/4 ). 
\end{eqnarray} 
Which means an approximate $1/\sqrt{sin(\theta)cos(\theta)}$ angle dependence which and has a strong decay for large scattering angles. 
  Note, that even this function shows very rapid oscillations. 
Figure \ref{negyes} shows the same kind of cross sections for for $z = 10000$ (which means $I= 4.56\times 10^{15}$ W/cm$^2$ intensity) 
and for  $z = 6.61\times 10^{6} $ (which means $I= 2.0\times 10^{21}$ W/cm$^2$ intensity), respectively. 
Only the $ n=1 $ one photon absorption process is considered.  


\section{Summary}
We presented a formalism which gives analytic angular differential cross section model for laser assisted proton 
nucleon scattering on a Woods-Saxon optical potential where the nth-order photon 
absorption is taken into account simultaneously. 
As an example the proton - $^{12}$Ca collision was investigated at moderate 49 MeV proton energies.
The laser intensities vary from from $10^7$ W/cm$^2$ to $  10^{21}$ W/cm$^2.$  
We found that at  $10^7$ W/cm$^2$ laser intensities the elastic Born total cross section is a factor of 90 times higher 
than the same process where a single photon is absorbed. At higher laser field intensities this ratio is even higher. 
We hope that this study will stimulate laser assisted nuclear scattering collisions which will be done in the Romanian ELI facility.   

\section{Acknowledgement} 
We thank for Prof. Gyula Bencze for useful discussions and comments.       
                                                                  
\end{document}